\documentclass[aps,prl,twocolumn,showpacs,superscriptaddress]{revtex4}
\usepackage{amsmath}
\usepackage{graphicx}
\usepackage{bm}
\usepackage{mathrsfs}
\begin{document}

\title{Generation of Gravitational Radiation in Dusty Plasmas and Supernovae}

\author{Gert Brodin} 
\altaffiliation[Also at: ]{Centre for Fundamental Physics,
 Rutherford Appleton Laboratory,
 Chilton Didcot, Oxfordshire, OX11 0QX, UK} 
\affiliation{Department of Physics, Ume{\aa} University,
SE--901 87 Ume{\aa}, Sweden}

\author{Mattias Marklund} 
\altaffiliation[Also at: ]{Centre for Fundamental Physics,
 Rutherford Appleton Laboratory,
 Chilton Didcot, Oxfordshire, OX11 0QX, UK} 
\affiliation{Department of Physics, Ume{\aa} University,
SE--901 87 Ume{\aa}, Sweden}

\author{Padma K.\ Shukla}
\altaffiliation[Also at: ]{Centre for Fundamental Physics,
 Rutherford Appleton Laboratory,
 Chilton Didcot, Oxfordshire, OX11 0QX, UK} 
\affiliation{Department of Physics, Ume{\aa} University,
SE--901 87 Ume{\aa}, Sweden} 
\affiliation{Institut f\"ur Theoretische Physik IV, Fakult\"at f\"ur
     Physik und Astronomie, Ruhr-Universit\"at Bochum, D--44780 Bochum,
     Germany}

\date{\today}

\begin{abstract}
We present a novel nonlinear mechanism for exciting a gravitational
radiation pulse (or a gravitational wave) by dust magnetohydrodynamic (DMHD)
waves in dusty astrophysical plasmas. We derive the relevant equations
governing the dynamics of nonlinearly coupled DMHD waves and a gravitational
wave (GW). The system of equations is used to investigate the generation of
a GW by compressional Alfv\'{e}n waves in a type II supernova. The growth
rate of our nonlinear process is estimated, and the results are discussed in
the context of the gravitational radiation accompanying supernova explosions.
\end{abstract}

\maketitle

%\title{Nonlinear Generation of Gravitational Radiation by Magnetohydrodynamic
%Waves in Dusty Astrophysical Plasmas}

It is well known that there exist numerous mechanisms for the conversion
between gravitational waves (GWs) and electromagnetic waves \cite
{grishchuk,Papadouplous2001,Moortgat2003,Moortgat2004,kallberg2004,%
Anile99,Servin2003,servinbrodin,Mendonca2002,Balakin2003,Li2002,%
BrodinMarklund2003,Picasso2003,Papadoupolus2002,MDB2000,%
Hogan2002,bms,Vlahos2004,Mosquera2002}. 
For example, the propagation of GWs across an external magnetic field
gives rise to a linear coupling to the electromagnetic field \cite{grishchuk}
, which may lead to the gravitational wave excitation of ordinary
electromagnetic waves in vacuum, or of magnetohydrodynamic (MHD) waves in a
plasma \cite{Papadouplous2001,Moortgat2003,Moortgat2004}. Furthermore,
various nonlinear coupling mechanisms give rise to three-wave couplings
between GWs and electromagnetic waves in matters. We also note that
four-wave processes may cause graviton-photon conversion even in the absence
of external matters or fields \cite{kallberg2004}. Moreover, GWs can couple
to other types of waves, e.g. sound waves, also in neutral media \cite
{Anile99}. There are numerous motives for considering wave couplings
involving GWs. In some cases, the emphasis is on the basic theory \cite
{kallberg2004,Anile99,Servin2003,servinbrodin,Mendonca2002,Balakin2003}. In
other works, the focus is on GW detectors \cite
{Li2002,BrodinMarklund2003,Picasso2003}, on cosmology \cite
{Papadoupolus2002,MDB2000,Hogan2002}, or on astrophysical applications such
as binary mergers \cite{bms}, gamma ray bursts \cite{Vlahos2004} or pulsars 
\cite{Mosquera2002}. Many of the previous works have concentrated on the
conversion from GWs to electromagnetic waves, which can be analysed within a
test matter approach which neglects the back reaction on the gravitational
field. We note that such an approach can be justified if the background
energy density is low.

In this Letter, we consider the three-wave coupling between two DMHD waves
and a GW, including the effects of dust particles \cite{shukla-mamun} in a
dense medium such as the supernova where electrons, protons, and charged
dust macroparticles are abundant. For this purpose, we derive the dust Hall
MHD equations \cite{shukla-mamun}, i.e. equations describing the dust MHD
waves, including the effect of a GW. We emphasize that for a low-beta
plasma, the system of equations has a structure which can describe both a
dust-dominated plasma, as well as an ordinary Hall-MHD plasma (if we replace
the dust mass density by the ion mass density). Using the normal mode
approach \cite{Weiland-Wilhelmsson}, the three-wave coupling equations are
derived, including the back reaction on a GW from the Einstein equations.
The system is shown to fulfil the Manley-Rowe relations \cite
{Weiland-Wilhelmsson} (which means that the interaction process can be
viewed quantum mechanically) and to be energy conserving. The three-wave
equations are then used to analyse the generation of a GW by the
compressional Alfv\'{e}n waves in the iron core of the type II supernova 
\cite{Raffelt,Woosley-etal}. It turns out that the characteristic timescale
for the Alfv\'{e}n-GW conversion can be less than a millisecond, which
implies that the mechanism is potentially relevant for the high-frequency
part ($> 1\mathrm{MHz}$) of the supernova GW spectrum.

The plasma dynamics, due to the response to a gravitational wave 
\begin{equation}
ds^{2}=-dt^{2}+\left( 1+h_{+}\right) dx^{2}+\left( 1-h_{+}\right)
dy^{2}+2h_{\times }dx\,dy+dz^{2}  \label{eq:gw}
\end{equation}
can be formulated according to 
\begin{equation}
\partial _{t}n_{s}+\nabla \cdot (n_{s}\mathbf{v}_{s})=0,
\label{eq:continuity}
\end{equation}
and 
\begin{equation}
m_{s}n_{s}(\partial _{t}+\mathbf{v}\cdot \nabla )\mathbf{v}=-\nabla
p_{s}+q_{s}n_{s}(\mathbf{E}+\mathbf{v}\times \mathbf{B})+m_{s}n_{s}\mathbf{g}%
_{s},  \label{eq:momentum}
\end{equation}
where we have introduced the tetrad $\mathbf{e}_{0}=\partial _{t},\quad 
\mathbf{e}_{1}=\left( 1- h_{+}/2 \right) \partial _{x} - h_{\times
}/2\partial _{y}, \mathbf{e}_{2}=\left( 1+ h_{+}/2\right) \partial _{y} -
h_{\times }/2\partial _{x}, \mathbf{e}_{3}=\partial _{z} $, and $\nabla =(%
\mathbf{e}_{1},\mathbf{e}_{2},\mathbf{e}_{3})$. Moreover, 
\begin{eqnarray}
\mathbf{g}_{s} &=&-\tfrac{1}{2}(1-v_{sz})(v_{sx}\partial
_{t}h_{+}+v_{sy}\partial _{t}h_{\times })\mathbf{e}_{1}  \notag \\
&& -\tfrac{1}{2}(1-v_{sz})(v_{sx}\partial _{t}h_{\times }-v_{sy}\partial
_{t}h_{+})\mathbf{e}_{2}  \notag \\
&&-\tfrac{1}{2}\left[ (v_{sx}^{2}-v_{sy}^{2})\partial
_{t}h_{+}+2v_{sx}v_{sy}\partial _{t}h_{\times }\right] \mathbf{e}_{3}
\end{eqnarray}
represents the gravitational acceleration of the particle species $s$ due to
the GWs. We have assumed that $\partial _{z}\approx -\partial _{t}$ holds
for the GWs.

The electromagnetic field is determined through the gravity modified
Maxwell's equations. Using the same notation as above, they take the form 
\begin{eqnarray}
\partial_{t}\mathbf{E} &=& \nabla\times\mathbf{B} - \sum_{s} q_{s}n_{s}%
\mathbf{v}_{s} - \mathbf{j}_{E} ,  \label{eq:ampere} \\
\partial_{t}\mathbf{B} &=& -\nabla\times\mathbf{E} - \mathbf{j}_{B} ,
\label{eq:faraday}
\end{eqnarray}
with the constraints $\nabla\cdot\mathbf{E} = \sum_{s} q_{s}n_{s} $ and $%
\nabla\cdot\mathbf{B} = 0$. Here, the effects of the GWs (\ref{eq:gw}) are
represented by the effective currents 
\begin{eqnarray}
\mathbf{j}_{E} &=& -\tfrac{1}{2}\left[ (E_{x} - B_{y})\partial_{t}h_{+}
+(E_{y} + B_{x})\partial_{t} h_{\times} \right]\mathbf{e}_{1}  \notag \\
&& -\tfrac{1}{2}\left[ -(E_{y} + B_{x})\partial_{t}h_{+} - (E_{x} -
B_{y})\partial_{t} h_{\times} \right]\mathbf{e}_{2} ,
\end{eqnarray}
and 
\begin{eqnarray}
\mathbf{j}_{B} &=& -\tfrac{1}{2}\left[ (E_{y} + B_{x})\partial_{t}h_{+} -
(E_{x} - B_{y})\partial_{t} h_{\times} \right]\mathbf{e}_{1}  \notag \\
&& -\tfrac{1}{2}\left[ (E_{x} - B_{y})\partial_{t}h_{+} +(E_{y} +
B_{x})\partial_{t} h_{\times} \right]\mathbf{e}_{2} .
\end{eqnarray}

With the general setting established above, we will from now on focus on the
case of a three-component dusty plasma, for which we have the equation of
state $%\begin{equation}
p_{s}=k_{B}T_{s}n_{s}$. %\end{equation}
Thus, the plasma is composed of electrons ($e$), ions ($i$), and dust
particles ($d$). The mass $m_{d}$ of the dust particles is assumed to be
much larger than the electron and ion masses, viz. $m_{e}$ and $m_{i}$,
respectively. We will assume that the plasma is approximately quasi-neutral,
i.e. $q_{i}n_{i}=en_{e}-q_{d}n_{d}%\label{eq:neutral}
$. Moreover, the waves under consideration are supposed to propagate with
phase velocities much smaller than the speed of light $c$. Thus, we may
neglect the displacement current in Amp\'{e}re's law (\ref{eq:ampere}), i.e. 
\begin{equation}
\nabla \times \mathbf{B}=\sum_{s}q_{s}n_{s}\mathbf{v}_{s}+\mathbf{j}_{E}.
\end{equation}
Due to the constraint $m_{e},m_{i}\ll m_{d}$ the momentum conservation
equation (\ref{eq:momentum}) for the inertialess electrons and ions becomes 
\begin{equation}
0=-k_{B}T_{e}\nabla n_{e}-en_{e}(\mathbf{E}+\mathbf{v}_{e}\times \mathbf{B}%
)+m_{e}n_{e}\mathbf{g}_{e},  \label{eq:momentum-e}
\end{equation}
and 
\begin{equation}
0=-k_{B}T_{i}\nabla n_{i}+q_{i}n_{i}(\mathbf{E}+\mathbf{v}_{i}\times \mathbf{%
B})+m_{i}n_{i}\mathbf{g}_{i},  \label{eq:momentum-i}
\end{equation}
respectively. Adding Eqs.\ (\ref{eq:momentum-e}) and (\ref{eq:momentum-i}),
using the quasineutrality condition, assuming that the number densities of
the electrons and ions are not much larger than the number density of the
dust, and using the heavy dust approximation, the dust momentum equation
takes the form 
\begin{eqnarray}
&&\rho _{d}\left( \partial _{t}+\mathbf{v}_{d}\cdot \nabla \right) \mathbf{v}%
_{d}=-k_{B}\left( T_{d}-\frac{q_{d}}{q_{i}}T_{i}\right) \nabla n_{d}  \notag
\\
&&\quad +(\nabla \times \mathbf{B})\times \mathbf{B}-\mathbf{j}_{E}\times 
\mathbf{B}+\rho _{d}\mathbf{g}_{d}  \label{eq:momentum-d}
\end{eqnarray}
where $\rho _{d}=m_{d}n_{d}$. In Eq.\ (\ref{eq:momentum-d}) we have also used
the approximation $[T_{e}+(e/q_{i})T_{i}]n_{e}\ll \lbrack
T_{d}-(q_{d}/q_{i})T_{i}]n_{d}$.

Again using Eqs.\ (\ref{eq:momentum-e}) and (\ref{eq:momentum-i}) to
eliminate the electric field, Faraday's law (\ref{eq:faraday}) becomes 
\begin{equation}
\partial_{t}\mathbf{B} = \nabla\times(\mathbf{v}_{d}\times\mathbf{B}) - 
\frac{m_{d}}{q_{d}}\nabla\times\left[ \left( \partial_{t} + \mathbf{v}%
_{d}\cdot\nabla \right)\mathbf{v}_{d} - \mathbf{g}_{d} \right] - \mathbf{j}%
_{B} ,  \label{eq:faraday2}
\end{equation}
where we have used the dust momentum equation (\ref{eq:momentum-d}).

Thus, Eqs.\ (\ref{eq:momentum-d}) and (\ref{eq:faraday2}) together with the
dust continuity equation 
\begin{equation}
\partial _{t}\rho _{d}+\nabla \cdot (\rho _{d}\mathbf{v}_{d})=0,
\label{Continuity}
\end{equation}
constitute the dust MHD equations in the presence of a GW. For a low-beta
plasma, the pressure term in (\ref{eq:momentum-d}) is negligible, which
means that the structure of Equations (\ref{eq:momentum-d})-(\ref{Continuity}%
) is the same as in an ordinary Hall-MHD plasma without dust. Henceforth, we
will consider a low-beta plasma, drop the index $d$ on all quantities and
thus let $q/m$ be either the ion charge to mass ratio or the considerably
smaller charge to mass ratio of the dust particles. As a result, our
mathematical analysis below will then apply either to a dust Hall-MHD
plasma, or to an ordinary Hall-MHD plasma without dust.

To simplify the problem, we consider the case when the dust-acoustic speed $%
c_{s}=k_{B}T/m$ is much smaller than the dust Alfv\'{e}n velocity $%
C_{A}=\left( B_{0}^{2}/\mu _{0}\rho \right) ^{1/2}$ such that the pressure
term in (\ref{eq:momentum-d}) can be neglected. As a prerequisite for the
nonlinear calculations, we first study the linear modes of the system (\ref
{eq:momentum-d})-(\ref{Continuity}) omitting the gravitational
contributions. Letting $B=B_{0}\widehat{\mathbf{z}}+\mathbf{B}_{1}$, $\rho =$
$\rho _{0}+\rho _{1}$, where the index $1$ denotes the perturbation of the
equilibrium part, and linearizing Eqs. (\ref{eq:momentum-d})-(\ref
{Continuity}) and Fourier analysing, we readily obtain the dispersion
relation 
\begin{equation}
\left( \omega ^{2}-k_{z}^{2}C_{A}^{2}\right) \left( \omega
^{2}-k^{2}C_{A}^{2}\right) -\frac{\omega ^{2}k_{z}^{2}k^{2}C_{A}^{2}}{\omega
_{c}^{2}}=0,  \label{Dispersion-relation}
\end{equation}
where $\omega _{c}=qB_{0}/m$ is the gyrofrequency. For frequencies much
smaller than the gyrofrequency, we note that the modes separate into the
shear Alfv\'{e}n wave, $\omega ^{2}-k_{z}^{2}C_{A}^{2}\approx 0$, and the
compressional Alfv\'{e}n wave, $\omega ^{2}-k^{2}C_{A}^{2}\approx 0$. Below
we will consider the more general case described by (\ref
{Dispersion-relation}), however. For later applications it is convenient to
use the linear equations to express all quantities in terms of a single
variable. Thus, we let the wavevector of the dust MHD waves lie in the $x-z$%
-plane, and write 
\begin{eqnarray}
v_{y} &=&i\frac{\omega }{\omega _{c}}\frac{k_{z}^{2}C_{A}^{2}}{\left( \omega
^{2}-k_{z}^{2}C_{A}^{2}\right) }v_{x},\,v_{z}=0,\,\rho _{1}=\rho _{0}\frac{%
k_{x}v_{x}}{\omega }  \label{help3} \\
B_{x} &=&-B_{0}\frac{\omega k_{z}}{k^{2}C_{A}^{2}}v_{x}  \label{help4} \\
B_{y} &=&-iB_{0}\frac{\omega ^{2}}{\omega _{c}}\frac{k_{z}}{\left( \omega
^{2}-k_{z}^{2}C_{A}^{2}\right) }v_{x}  \label{help5} \\
B_{z} &=&B_{0}\frac{\omega k_{x}}{k^{2}C_{A}^{2}}v_{x}.  \label{help6}
\end{eqnarray}

Next, we consider a system of three weakly interacting waves. Two dust MHD
waves with frequencies and wave-numbers $\left( \omega _{1},\mathbf{k}%
_{1}\right) $ and $\left( \omega _{2},\mathbf{k}_{2}\right) $ respectively,
and an arbitrarily polarized gravitational wave propagating along the
z-direction with the frequency and wavenumber $\left( \omega _{g},k_{g}%
\widehat{\mathbf{z}}\right) $. Noting that the gravitational dispersion
relation reads $\omega _{g}=k_{g}c$ and that $C_{A}\ll c$, the frequency and
wavenumber matchings (energy and momentum conservation) can be approximated 
\begin{eqnarray}
\omega _{g} =\omega _{1}+\omega _{2}, \, \mathbf{k}_{g} =\mathbf{k}_{1}+%
\mathbf{k}_{2}\Rightarrow 0 \approx \mathbf{k}_{1}+\mathbf{k}_{2}.
\label{Energy-match}
\end{eqnarray}
We will thus use $\mathbf{k}_{1}\approx -\mathbf{k}_{2}$ below, and we
define $k_{1x}=-k_{2x}\equiv k_{x}$ as well as $k_{1z}=-k_{2z}\equiv k_{z}$
(letting $k_{y}=0$ for convenience). All quantities are now assumed to be
superpositions of two dust MHD waves waves whose amplitudes are weakly
varying functions of time, i.e. $\rho =\rho _{0}+\sum_{j=1}^{2}\rho
_{j}(t)\exp [i(\mathbf{k}_{j}\cdot \mathbf{r-\omega t})]+\mathrm{c.c.}$,
where $\mathrm{c.c}$ stands for the complex conjugate, In principle, the
gravitational wave should also contribute, but we note that within a fluid
model the gravitational wave contribution to all plasma perturbations
(velocity, magnetic field and density) are second order in the gravitational
wave amplitudes, provided that the GW propagates parallel to $\mathbf{B}_{0}$%
, as we have assumed. Thus, the only linear perturbations due to the
gravitational wave are those of the metric as described by Eq. (\ref{eq:gw}).

Next, in order to simplify the algebra, we introduce the normal mode $a_{j}$
defined by 
\begin{eqnarray}
&&\!\!\!\!\!\!a_{j}=\frac{\omega _{j}}{k_{xj}}v_{xj}-i\frac{\omega
_{j}^{2}k_{zj}^{2}C_{A}^{2}}{\omega _{c}\left( \omega
_{j}^{2}-k_{zj}^{2}C_{A}^{2}\right) k_{xj}}v_{yj}-\frac{\omega _{j}^{2}}{%
k_{j}^{2}}\frac{k_{zj}}{k_{xj}B_{0}}B_{xj}  \notag \\
&&\!\!\!\!\!\!+\frac{\omega _{j}^{2}k_{z_{j}}^{2}C_{A}^{2}}{\omega
_{c}^{2}\left( \omega _{j}^{2}-k_{zj}^{2}C_{A}^{2}\right) }\frac{C_{A}^{2}}{%
B_{0}}B_{z}  \notag \\
&&+i\frac{\left( \omega _{j}^{2}-k_{j}^{2}C_{A}^{2}\right) \omega
_{c}^{2}-k_{j}^{2}C_{A}^{2}\omega _{j}^{2}}{k_{j}^{2}\omega _{j}}\frac{%
k_{z_{j}}}{\omega _{c}k_{xj}B_{0}}B_{y}.  \label{Normal mode}
\end{eqnarray}
Returning to Eqs. (\ref{eq:momentum-d})-(\ref{Continuity}) and including the
nonlinear terms \cite{Tetrad-note}, we can, by keeping the part varying as $%
\exp [i(\mathbf{k}_{1}\cdot \mathbf{r-\omega }_{1}\mathbf{t})]$, derive 
\begin{eqnarray}
&&\!\!\!\!\!\!\!\!\!\!\frac{\partial a_{1}}{\partial t}=-\frac{\omega _{1}}{%
2k_{x}}\left( v_{2x}^{\ast }h_{+}+v_{2y}^{\ast }h_{\times }\right) -i\frac{%
\omega _{1}^{2}k_{z}^{2}C_{A}^{2}\left( v_{2x}^{\ast }h_{\times
}-v_{2y}^{\ast }h_{+}\right) }{2\omega _{c}\left( \omega
_{1}^{2}-k_{z}^{2}C_{A}^{2}\right) k_{x}}  \notag \\
&&\!\!\!\!\!\!\!\!\!\!-\frac{\omega _{1}^{2}}{2k^{2}}\frac{k_{z}}{k_{x}B_{0}}%
\left( B_{2x}^{\ast }h_{+}-B_{2y}^{\ast }h_{\times }\right)   \notag \\
&&\!\!\!\!\!\!\!\!\!\! + i\frac{k_z\left[\left( \omega _{1}^{2}-k^{2}C_{A}^{2}\right)
\omega _{c}^{2}-k^{2}C_{A}^{2}\omega _{1}^{2}\right]}{k^{2}\omega _{1}\omega _{c}k_{x}B_{0}}{%
}\left( B_{2x}^{\ast }h_{\times }-B_{2y}^{\ast
}h_{+}\right) ,  \label{Intermediate}
\end{eqnarray}
and we obtain a similar result for $\partial a_{2}/\partial t$ by letting $%
1\leftrightarrow 2$. After some algebra, using Eqs. (\ref{help3})-(\ref
{help6}) and (\ref{Normal mode}), we find that Eq. (\ref{Intermediate})
reduces to 
\begin{equation}
\frac{dv_{x1}}{dt}=\frac{\rho _{0}\omega _{1}}{W_{1}}v_{x2}^{\ast }\left(
V_{+}h_{+}+V_{\times }h_{\times }\right) ,  \label{Coupling1}
\end{equation}
and similarly for mode 2 
\begin{equation}
\frac{dv_{x2}}{dt}=\frac{\rho _{0}\omega _{2}}{W_{2}}v_{x1}^{\ast }\left(
V_{+}h_{+}+V_{\times }h_{\times }\right) ,  \label{Coupling2}
\end{equation}
where 
\begin{eqnarray}
&&W_{1,2}=\frac{\rho _{0}}{2}\Bigg[1+\frac{\omega _{1,2}^{2}}{%
k_{1,2}^{2}C_{A}^{2}}  \notag \\
&&\quad +\frac{\omega _{1,2}^{2}k_{z1,2}^{4}C_{A}^{2}}{\omega _{c}^{2}\left(
\omega _{1,2}^{2}-k_{z1,2}^{2}C_{A}^{2}\right) ^{2}}\left( 1+\frac{\omega
_{1,2}^{2}}{k_{z1,2}^{2}C_{A}^{2}}\right) \Bigg]  \label{Energy} \\
&&V_{\times }=i\left[ \frac{\omega _{1}k_{z}^{2}C_{A}^{2}k^{2}-\omega
_{2}\omega _{1}^{2}k_{z}^{2}}{\omega _{c}\left( \omega
_{1}^{2}-k_{z1}^{2}C_{A}^{2}\right) k^{2}}+1\leftrightarrow 2\right] ,
\label{Coeff+}
\end{eqnarray}
and 
\begin{equation}
V_{+}=i\left[ 1+\frac{\omega _{1}\omega _{2}}{\omega _{c}^{2}}\frac{%
k_{z}^{2}C_{A}^{2}\left( k_{z}^{2}C_{A}^{2}+\omega _{1}\omega _{2}\right) }{%
\left( \omega _{1}^{2}-k_{z}^{2}C_{A}^{2}\right) \left( \omega
_{2}^{2}-k_{z}^{2}C_{A}^{2}\right) }+\frac{\omega _{1}\omega _{2}k_{z}^{2}}{%
k^{4}C_{A}^{2}}\right] .  \label{Coeff-x}
\end{equation}
Next using the Einstein equations, linearized in $h_{+},h_{\times }$,
keeping only the resonantly varying part of $T^{\mu \nu }$ we obtain for the 
$\times $ and $+$-polarization, respectively 
\begin{equation}
i\omega _{g}\frac{dh_{\times }}{dt}=\kappa \left[ \rho _{0}v_{x1}v_{y2}+%
\frac{B_{x2}B_{y1}}{\mu _{0}}\right] ,  \label{x-help}
\end{equation}
and 
\begin{equation}
i\omega _{g}\frac{dh_{\times }}{dt}=\frac{\kappa }{2}\left[ \rho _{0}\left(
v_{x1}v_{x2}-v_{y1}v_{y2}\right) +\frac{B_{x1}B_{x2}-B_{y1}B_{y2}}{\mu _{0}}%
\right] ,  \label{+-help}
\end{equation}
which is reduced to 
\begin{equation}
\frac{dh_{\times }}{dt}=-\frac{\rho _{0}\omega _{g}}{W_{g}}V_{\times
}v_{x1}v_{x2},  \label{x-result}
\end{equation}
and 
\begin{equation}
\frac{dh_{+}}{dt}=-\frac{\rho _{0}\omega _{g}}{W_{g}}V_{+}v_{x1}v_{x2},
\label{+-result}
\end{equation}
where $%\begin{equation}
W_{g}={\omega _{g}^{2}}/{2\kappa }% \label{GW-energy}
$. %\end{equation}
The total wave energy is $W_{\mathrm{tot}}=W_{1}\left| v_{x1}\right|
^{2}+W_{2}\left| v_{x2}\right| ^{2}+W_{g}(\left| h_{\times }\right|
^{2}+\left| h_{+}\right| ^{2})$ %\begin{equation}
%W_{\mathrm{tot}}=W_{1}\left| v_{x1}\right| ^{2}+W_{2}\left| v_{x2}\right|
%^{2}+W_{g}\left( \left| h_{\times }\right| ^{2}+\left| h_{+}\right|
%^{2}\right),   \label{Wtot}
%\end{equation}
and it is easily verified from (\ref{Coupling1})-(\ref{Coupling2}) together
with (\ref{x-result})-(\ref{+-result}) that $W_{\mathrm{tot}}$ is conserved.
Furthermore, the appearance of the same coupling coefficients $V_{\times }$, 
$V_{+}$ in Eqs. (\ref{Coupling1})-(\ref{Coupling2}) as well as in (\ref
{x-result})-(\ref{+-result}) assures that the Manley-Rowe relations are
fulfilled, which implies that each mode changes energy in direct proportion
to its frequency, i.e. $(dW_{1}/dt)/(dW_{2}/dt)=\omega _{1}/\omega _{2}$
etc. The system of (\ref{Coupling1})-(\ref{Coupling2}) together with (\ref
{x-result})-(\ref{+-result}) describing the energy conversion between DMHD
waves and GWs is one of the main results of the present letter. A more
elaborate calculation scheme, including effects such as inhomogeneity and
background curvature, is a project for future research.

We now apply our results to the gravitational radiation arising from the
iron core of a type II supernova where the densities can be of the order $%
10^{17}\,\mathrm{kg/m}^{3}$ \cite{Woosley-etal}. The large neutrino outflow
(which can reach powers of $10^{33}\,\mathrm{W/cm}^{2}$ see e.g. Ref. \cite
{Raffelt}) can generate MHD waves described by (\ref{Dispersion-relation}).
From the flux conservation, we expect the iron core to be strongly
magnetized (comparable to pulsars), and for magnetic field strengths $%
B_{0}\sim 10^{8}\,\mathrm{T}$, the gyrofrequency will be much larger than
all other frequencies of the problem. The dispersion relation then separate
into the shear Alfv\'{e}n waves $\omega ^{2}-k_{z}^{2}C_{A}^{2} \approx 0$
and the compressional Alfv\'{e}n waves $\omega ^{2}-k^{2}C_{A}^{2} \approx 0$. 
Assuming that the pump MHD wave is a compressional mode with a
frequency $5\,\mathrm{MHz}$, the matching conditions (\ref{Energy-match}) 
%and (\ref{vector-match})
can be fulfilled for a GW with a typical frequency $3\mathrm{MHz}$ and a
shear Alfv \'{e}n wave with the frequency $2\,\mathrm{MHz}$ \cite
{Matching-note}. For the assumed geometry, the MHD waves couple only to the $%
h_{\times }$-polarization (to a good approximation), and by combining (\ref
{Coupling2}) and (\ref{x-result}), we obtain 
\begin{equation}
\frac{d^{2}h_{\times }}{dt^{2}}=-h_{\times }\frac{\omega _{2}}{\omega _{g}}%
\frac{\left| V_{\times }\right| ^{2}}{W_{2}}\frac{16\pi G}{c^{2}}\rho
_{0}\left| v_{x1}\right| ^{2}.  \label{Growth}
\end{equation}
Thus, noting that the factor $\omega _{2}\left| V_{\times }\right|
^{2}/\omega _{g}W_{2}$ is negative \cite{Matching-note} and has a
magnitude of order unity \cite{Unity-note} for the given parameters, 
the growth rate is 
\begin{equation}
\gamma \sim \sqrt{16\pi G\rho _{0}}\frac{\left| v_{x1}\right| }{c},
\label{Growth-estimate}
\end{equation}
which for a weakly relativistic pump quiver speed, $\left| v_{x1}\right|
/c\sim 1/10$, implies $\gamma \sim 10\,\mathrm{kHz}$. Thus, we deduce that
excitation of a GW by MHD waves is a reasonably fast process in a dense
medium such as the supernova iron core. Furthermore, the present process can
give rise to GWs of higher frequencies than many of the previously
considered excitation mechanisms, see e.g. \cite{Cutler-Thorne}. Thus, our
model contributes to the understanding of gravitational radiation 
emissions accompanying supernova explosions \cite{sazhin}.


\begin{thebibliography}{99}
\bibitem{Tetrad-note}  When including the nonlinear GW-coupling, it is in
principle important to separate the coordinate components (indices $x,y,z$)
from the tetrad components (indices $1,2,3$), since the difference is first
order in $h_{+},h_{\times }$. However, for notational convenience we let
indices $x,y,z$ denote tetrad components $1,2,3$ in equation 
(\ref{Intermediate}) and henceforth.
\bibitem{Matching-note}  Note that from the Manley-Rowe relations the pump
wave must have the highest frequency for the growth rates to become real and
positive, Thus using the matching condition (\ref{Energy-match}) we see that
the shear Alfv\'{e}n mode formally has a negative frequency. Furthermore, we
note that the dispersion relations together with the matching conditions
means that the pump mode cannot propagate purely along the magnetic field.
Specifically, the wavevector matching implies $k_xC_A \approx 4.6\mathrm{%
MHZ}$, where $C_{A}\approx 300\mathrm{m/s}$ for the given parameters.
\bibitem{Unity-note}  This can be seen by taking the limit $\omega
_{c}\rightarrow \infty $ of the expression, noting that $\left( \omega
_{2}^{2}-k_{z2}^{2}C_{A}^{2}\right) $ scales as $1/\omega _{c}^{2}$ from the
dispersion relation (\ref{Dispersion-relation}).
\end{thebibliography}
\end{document}